\documentclass[12pt,preprint]{aastex}
\usepackage{graphicx}
\shorttitle{}
\shortauthors{}
\makeatletter
\newcommand{\thereaction}{\thechapter-\Roman{reaction}}
\@addtoreset{reaction}{chapter}
\newcounter{reaction}
\newdimen\reactionindent
\newdimen\reactionarrspc
\def\@reactionnum{\hbox{\reset@font\rm(\thereaction)}}
\newcommand\chem[1]{$\rm #1$}
{%
  \@beginparpenalty\predisplaypenalty%
  \@endparpenalty\postdisplaypenalty%
  \refstepcounter{reaction}%
  \trivlist \item[]\leavevmode%
    \hb@xt@\linewidth\bgroup $\m@th% $
    \displaystyle%
    \hskip\reactionindent%
    \rm
}{%
    $\hfil % $%
    \displaywidth\linewidth\hbox{\@reactionnum}%
  \egroup%
  \endtrivlist%
}

\newcommand\reactionlabel[1]{%
  $\m@th$\hfil%
  \refstepcounter{reaction}%
  \hbox{\@reactionnnum}%
  \label{#1}
}
\newdimen\autotop  \newdimen\autobottom  \newdimen\autosize 
\makeatother
\begin{document}
\bibliographystyle{apj}
\title{Turbulence driven diffusion in protoplanetary disks -- chemical effects
in the outer regions}
\author{Karen Willacy\altaffilmark{1}, William Langer\altaffilmark{1}, Mark Allen\altaffilmark{1,2} \& Geoffrey Bryden\altaffilmark{1}}
\altaffiltext{1}{Jet Propulsion Laboratory, California Institute of Technology, 
Pasadena CA 91109}
\altaffiltext{2}{Division of Geological Sciences, California Institute of Technology,
Pasadena, CA 91125}
\email{Karen.Willacy@jpl.nasa.gov}

\begin{abstract}
The chemistry of the early stages of disk evolution can be used observationally
to trace their physical characteristics, but it also determines their
later development, including the type and composition of any planets that
may form.
Modeling can be used in conjunction with molecular observations
to understand the extent to which
interstellar material incorporated into disks has
been processed and can elucidate the nature of the
processes acting in the disks.  One aspect that has
not received much attention until recently is the effect
of mixing processes on the observed chemical abundances.
The dynamics and chemistry of disks are likely to be intricately linked,
with dynamical processes altering the chemical composition and
chemistry, in turn, controlling the ionization structure and hence the ability 
of the magneto-rotational instability to drive disk turbulence.
Here we present the results from the first chemical
models of the outer regions ($R$ $>$ 100 AU) of
protoplanetary disks to consider the effects of turbulence
driven diffusive mixing in the vertical direction.
We concentrate 
on the outer disk to facilitate comparison with current millimeter and
sub--millimeter observations.
We show that vertical diffusion can greatly affect the column densities
of many species, increasing them by factors of up to two orders of
magnitude.  
Previous disk models have shown that disks can be divided into
three chemically distinct layers, with the bulk of the observed
molecular emission coming from a region
between an atomic/ionic layer on the surface of the
disk and the midplane region where the bulk of molecules
are frozen onto grains.  Diffusion retains this three layer structure, but
increases the
depth of the molecular layer by bringing atoms and atomic ions
formed by photodissociation in the surface layers into the shielded
molecular layer where molecules can reform.  For other species, notably
\chem{NH_3} and \chem{N_2H^+}, the column densities are relatively unaffected by diffusion.
These species peak in abundance in the midplane, where most other molecules
are heavily depleted, rather than in the molecular layer above.  Diffusion
only affects the abundances of those molecules with peak abundances
in the molecular layer.  We find that diffusion does not
affect the ionization fraction of the disk.
We compare the calculated
column densities to observations of DM Tau, LkCa 15 and TW Hya and find that
good agreement for many molecules with a diffusion coefficient of 10$^{18}$ cm$^2$s$^{-1}$.

\end{abstract}


\section{Introduction}

The collapse of a molecular cloud core to form a star results in the
formation of a disk of gas and dust around the young stellar object
(YSO) in its center.  
Observations from the millimeter and infrared probe the gas
and dust in the disks and are a primary source of information about
their characteristics, providing compositional information as well as
details of the density and temperature structure.  Chemical processing
in the disk affects the molecular abundances and hence what can be
observed.  The early stages of the disk evolution affect the
composition and evolution of the later, planet--forming, stages by
determining the nature of the material that is available for
coagulation into planets.  Understanding this stage of the star and
disk formation process is therefore crucial for the understanding of
the development of bodies such as are found in our own Solar System.

Most models to date have assumed that the disk is static,
e.g.\ \citet{wl00,aikawa02,wlb06}, or is accreting radially towards the
star
\citep{duschl96,bauer97,fg97,fgd97,willacy98,aikawa99,ah99,markwick02}.
One process that has not received much attention until recently is how
molecular abundances can be affected by mixing processes in the disk.
Previous models that have considered such processes, in the radial
direction only, were developed by \citet{morfill83, mv84, stevenson90,
gail01, wg02}.  The first study of mixing in the vertical direction
was carried out by \citet{ilgner04}.  They found that this process
changes the global chemical evolution of the inner 10 AU of a
protoplanetary disk.  In the protosolar nebula, mixing processes have
been shown to be potentially important and may provide a possible
explanation of the origin of crystalline silicates in comets
\citep{hanner99} and influence the degree of molecular deuteration
\citep{drouart99}.  Considering that protostellar disks are thought to
be highly turbulent, with the turbulence influencing their physical
evolution, it seems likely that this could also influence their
chemical evolution by providing a mechanism whereby molecules from
different parts of the disk can be mixed, driving chemical reactions,
and therefore altering the abundances of observable molecules.

Turbulence is a 3-D process and a complete treatment of turbulent
driven diffusive mixing would require the inclusion of a full chemical
network in a 3-D MHD code.  This approach is computationally
prohibitive at the present time.  Instead insight into the interplay
of diffusion and chemistry can be gained by considering a simplified
description of the dynamics which allows for a complicated chemical
network.  Here we present results of 1-D diffusive vertical mixing
models in the outer regions ($R$ $>$ 100 AU) of a protostellar disk.
We chose to study the outer disk because most observations with
current instrumentation are sensitive to this region.  Our work is
therefore relevant to the interpretation of observations of gaseous
molecules in T Tauri disks.  We consider how the magnitude of the
diffusion coefficient affects the predicted column densities and
demonstrate that the results are molecule dependent.  We find that
diffusion is potentially very important, blurring the distinction
between the different chemical layers in the disk, and that it will
have observable consequences.

\section{The diffusive mixing model}

Star formation begins with the collapse of a rotating molecular
cloud core.  This process is very rapid, too rapid to allow the
dissipation of angular momentum, since the freefall time
is shorter than, or comparable to, the angular momentum
transport time \citep{shu95}. Collapse therefore often
results in either the formation of a system of multiple
stars, or of a disk around the young star, or both.  Observations
have shown that disks
are extremely common around young stars.  Surveys of the Taurus
region indicate that roughly 50\% of pre-main sequence stars
have disks at an age of 1 Myrs \citep{strom89, beckwith90,
skrutskie90, strom93}.

Although the evolution of the star--disk system is controlled
by angular momentum transport, the mechanism by which this
transport is achieved is still unclear.  Molecular viscosity is
too low to have much of an effect.  Partly because
disks have a high Reynolds number (much higher
than is required in hydrodynamic flows for turbulence to set in),
turbulence is thought to be the means by which disks can rid themselves
of angular momentum (see \citet{mv84} and references
therein).  Several processes have been suggested to 
drive the turbulence,
e.g.\ the vertical convective instability \citep{lp80}, 
the magneto--rotational instability (MRI) \citep{bh91, bhs96},
the linear Rossby wave instability \citep{li00}, the baroclinic instability
\citep{kb03, klahr04} and the linear strato--rotational instability 
\citep{dubrulle05, sr05}.  The specific 
turbulent mechanism
is important because it is linked to the physical properties of the
disk such as its vertical structure, magnetization, temperature
and turbulent velocity dispersion.

Whatever the driving mechanism, a simple method to parameterize
the viscous stress in a disk is the $\alpha$ prescription
of \citet{ss73}.  This describes the angular momentum
transport in accretion disks in terms of an 'anomalous' viscosity,
$\nu_t$, proportional to the gas pressure:
\begin{equation}
\nu_t = \alpha c_s h
\end{equation}
where $c_s$ is the sound speed, $h$ is the vertical pressure scale height
and $\alpha$ is a dimensionless constant (where $\alpha$ $<$ 1).
The $\alpha$ parameter controls the viscous heating, angular momentum
transport and mixing in the disk.  Simulations using this prescription
can provide the physical parameters (density, temperature, vertical
structure) of a disk e.g.\ \citet{dalessio99,dalessio01,bryden06}.  
Estimates from the lifetimes of 
protostellar disks suggest that $\alpha$ $\sim$ 10$^{-2}$ \citep{chs00}, 
but model values between 10$^{-4}$ and 10$^{-1}$ are commonly used.

\begin{figure}
\epsscale{0.5}
\plotone{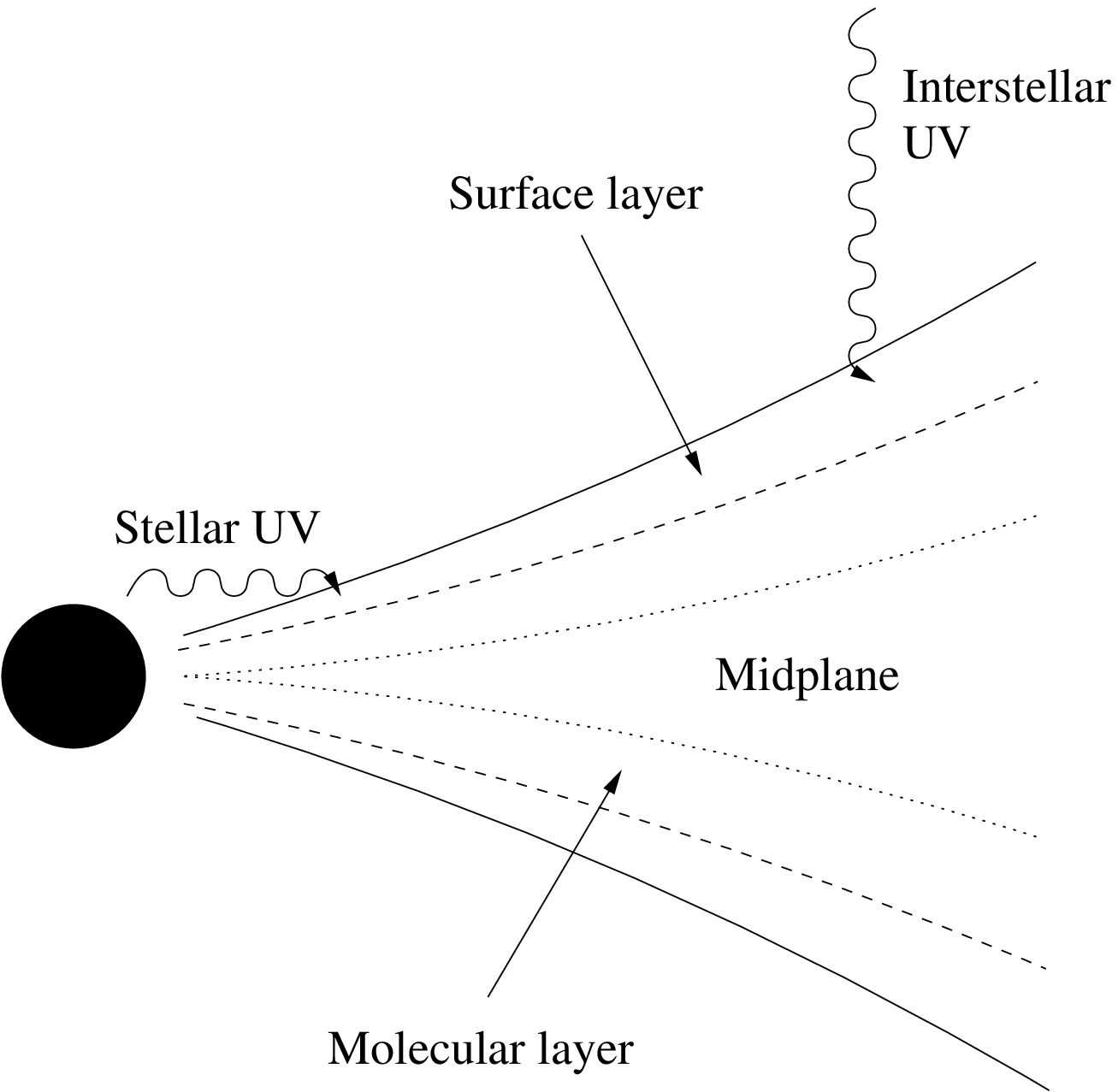}
\caption{\label{fig:disk_chem}Schematic of the chemical structure
of a protostellar disk showing how the disk can be divided up into
three chemically distinct layers.}
\end{figure}

The $\alpha$--disk models have been used to provide the density and
temperature structure for many of the current disk chemical models.
Various assumptions 
in the physical parameters have led to quantitative differences
in the results from these models.  Temperature is especially important
as it controls the location of the snow--line -- the region
where molecules are released from icy grain mantles by thermal
desorption. 
In addition, the chemical evolution is determined by
processes such as gas phase two- or three-body reactions,
cosmic ray ionization, X-ray and UV irradiation, gas--grain interactions
and grain surface reactions.  
 Qualitatively the models are in agreement with 
a three-layer structure being predicted (Figure~\ref{fig:disk_chem}).  
In the cold midplane
molecules are mainly accreted onto the dust grains, with little remaining
in the gas apart from H$_2$ and its ions.  At the surface of the disk,
the gas is warmed by irradiation from the central star.  Here molecules
can be thermally desorbed from the mantles and, once in the
gas, are quickly photodissociated by the strong UV field from
both the interstellar medium and from the star.  Between these
two layers is a region which is shielded from the external UV 
but where molecules can be kept in the gas and prevented from 
complete freezeout by desorption processes, either thermal
or non--thermal.  There is still 
a great deal of debate as to which non--thermal processes might
be acting.  Here we include thermal desorption and cosmic ray heating
of grains but ignore other non-thermal processes in order to concentrate
on the effects of diffusion alone.  

The diffusive mixing model used here was initially developed to model the
chemistry of planetary atmospheres \citep{ayw81}.
It also has been used by us to model molecular cloud
chemistry \citep{xal95,wla02}.
The method assumes that the turbulent diffusion timescale for a
given tracer depends on its composition gradient.  If the disk is
divided up into a series of zones, then diffusion will only occur
between two adjacent zones if there is a composition gradient between
them.  If the transport timescale between the zones is less than
the chemical timescale then the chemistry will be coupled to the turbulent
diffusion.

In 1-D the problem can be parameterized as follows \citep{ayw81}.
We assume that $n$(H$_2$) is the number density of H$_2$, $n_i$ is
the number density of molecule $i$ and $x_i$ is the fractional
abundance of $i$ (= $n_i/n$(H$_2$)).  Using mixing length theory
(which characterizes turbulence as eddies that maintain different
properties to the average fluid in which they exist for the time
taken to travel a distance $l$ - known as the mixing length), we 
can approximate fluctuations in the fractional abundance, $\delta x_i$,
due to the turbulence as
\begin{equation}
\delta x_i = -l \frac{dx_i}{dz}
\end{equation}
where $dx_i/dz$ is the abundance gradient in the $z$-direction.  The
net transport flux of $i$ is then given by
\begin{eqnarray}
\phi_i & = & n(\hbox{H}_2) < v_t \delta x_i > \nonumber \\
 & = & -K n(\hbox{H}_2)\frac{dx_i}{dz} \nonumber \\
 & = & -K n_i \left( \frac{1}{n_i}\frac{dn_i}{dz} - \frac{1}{n(\hbox{H}_2)}\frac{dn(\hbox{H}_2)}{dz}
\right)
\end{eqnarray}
where $K$ is the diffusion coefficient, defined as $K$ = $<v_t l>$ and $v_t$ is
the turbulent velocity.  Using this description of 1-D diffusion, we can 
write the chemical continuity equations as
\begin{equation}
\frac{\partial n_i}{\partial t} + \frac{\partial \phi}{\partial z} = P_i - L_i
\end{equation}
where $P_i$ and $L_i$ are the chemical production and loss terms for
species $i$, respectively.  
%The diffusion code solves the coupled continuity
%equations for all species simultaneously using an implicit finite difference
%representation adapted from \citet{richtmeyer}.
The diffusion--chemistry code solves the coupled continuity equations for
all species simultaneously by first linearizing it around an initial guess.
The subsequent linear parabolic partial differential equations are solved
using a completely implicit finite difference method described by
\cite{richtmeyer} (pp101-105, this scheme is not in the later edition).  It
can be shown that the numerical solutions converge as ($\Delta$t) and
($\Delta$z)$^2$.  The solution is iterated until the absolute fractional
errors are less than 10$^{-4}$.  A similar method was used by \cite{se70}
for modeling the F2 layer in the terrestrial atmosphere.  In this paper calculations
are carried out with a vertical spacing of 2 AU, so the number of zones varies
with radius.
%Calculations are carried out
%with a vertical spacing of 2 AU, so the number of zones varies
%with radius.

We assume that there is no infall into the disk, an assumption that
is appropriate for the T Tauri phase where
accretion is of less importance than in the earlier stages of 
the star formation process.  We also assume that material cannot
escape through the outer boundary, so the total flux across
it is zero.  We assume that the other boundary is the midplane,
and again no flux crosses this boundary.

\subsection{Determining the diffusion coefficient}
A critical factor in our model is the choice of diffusion coefficient, $K$.
We can define reasonable values of the diffusion coefficients if the
mixing process responsible for the diffusion of chemical species is also
responsible for the dispersion of angular momentum.  In this case we
can estimate the diffusion coefficient from the viscosity and
hence $K$ =
$\nu_t$ = $\alpha c_s h$.
This approach has been used previously by several authors,
e.g.\ \citet{morfill83, mv84, stevenson90,
gail01, wg02} to model the radial transport
in disks.  

We assume that the gas and dust are well mixed and that
the dust grains are small enough that they experience
mixing with the same diffusion coefficient as the gas.
Grains whose stopping times are less than the orbital
period are likely to diffuse with the gas.  Under
the conditions in our disk this means that grains with
radii up to several tens of centimeters will move with
the gas.
The transport of
grains from midplane to the heated surface regions means that
desorption from grains can potentially provide a source of
molecules, atoms and ions in the upper regions of the disk.
This will be discussed in our results below.  One 
limitation of our model is that the mixing of a given species 
is calculated based on its concentration gradient.  Thus 
the mantle species, rather than the grain itself, are assumed to
diffuse.  This will be addressed in future work.

Recent work by \citet{jk05} has measured directly the turbulent
diffusion coefficient of dust grains embedded in an MHD disk model
and compared this to the turbulent viscosity of the flow.
They found that the Schmidt number ($S_C$ = $\nu_t/K$) is about 1.5
in the vertical direction and 0.8 radially.  However
these results do not agree with those of \citet{csp05}  who
found $K$ $\sim$ 0.1 $\nu_t$.  More work is required to determine
the correct value.  In the meantime, we have elected to assume that 
$S_C$ = 1, i.e.\ $K$ = $\nu_t$.  

Assuming $K$ = $\nu_t$ we can find a typical value of $K$ in 
our disk. At a radial distance of 100 AU from the star, our disk model has a
gas temperature of 75 K at the surface giving $c_s$ = 4.4 $\times$ 10$^4$
cms$^{-1}$.  Taking $h/R$ = 0.15 and $\alpha$ = 10$^{-2}$ we find
$K$ = 9.9 $\times$ 10$^{16}$ cm$^2$s$^{-1}$.  Because
of the uncertainties in estimating $K$, and because changing
$\alpha$ could alter $K$ by orders of magnitude, (disk models
use values of $\alpha$ between 10$^{-4}$ and 10$^{-1}$), we have chosen
to test the effects of diffusion on chemistry by considering
values of $K$ between 0 and 10$^{18}$ cm$^2$s$^{-1}$.
We also make the simplifying assumption that $K$ does not vary
with position in the disk.  This spread
of diffusion coefficients will be sufficient to show us whether
or not diffusion affects the chemistry in the outer
disk and how strong the mixing needs to be to have observable
consequences.  

\section{The disk model}

The disk model used here is that of \citet{bryden06}.  This model describes
the structure of a geometrically thin, flared, 
disk which is heated both viscously
and by radiation from the central star.  The model assumes that the disk
is in steady state, i.e.\ that the disk properties are independent of time
and that there is a constant mass accretion rate.  The structure of
the disk is determined by solving the hydrodynamical equations 
semi--analytically.

The model we use has a mass accretion rate = 10$^{-8}$ $M_\odot$ yr$^{-1}$.
The central star has a mass of 0.7 $M_\odot$, temperature $T_*$ = 4000 K,
and radius $R_*$ = 2.4 $R_\odot$.
The parameter $\alpha$ is set at  0.01.  The surface density is $\Sigma_0$ = 
1000 gcm$^{-2}$ at 1 AU
and the density varies with radius as $\rho(R)$ $\propto$ $R^{-3/2}$ for 
$R$ $>$ 16 AU.  The bulk opacity of the disk has two values only: one
to visible starlight ($\kappa$ = 0 cm$^2$g$^{-1}$ in the upper atmosphere)
and one to its own infrared radiation ($\kappa$ = 2 $\times$ 10$^{-4}$ $T^2$
cm$^2$s$^{-1}$ within the disk).  This is a similar approach to that
taken by \cite{cg97}.  The grain size distribution is assumed
to follow an equilibrium cascade power--law, $da/dn$ $\propto$ $a^{-3.5}$,
such that the mass is concentrated in the largest particles, while the area
(and hence the opacity) is dominated by the smallest particles.  This
more closely resembles a Pollack distribution \citep{pollack94},
rather than a small--grain interstellar population.  The mean molecular
weight is 2.35.  The disk structure is calculated between radii of 0.05 and
500 AU and from 0 to $\pi$/4 in meridional angle.  The
total disk mass in this domain is $\sim$ 0.05 $M_\odot$.

\begin{figure}
\epsscale{1.0}
\plottwo{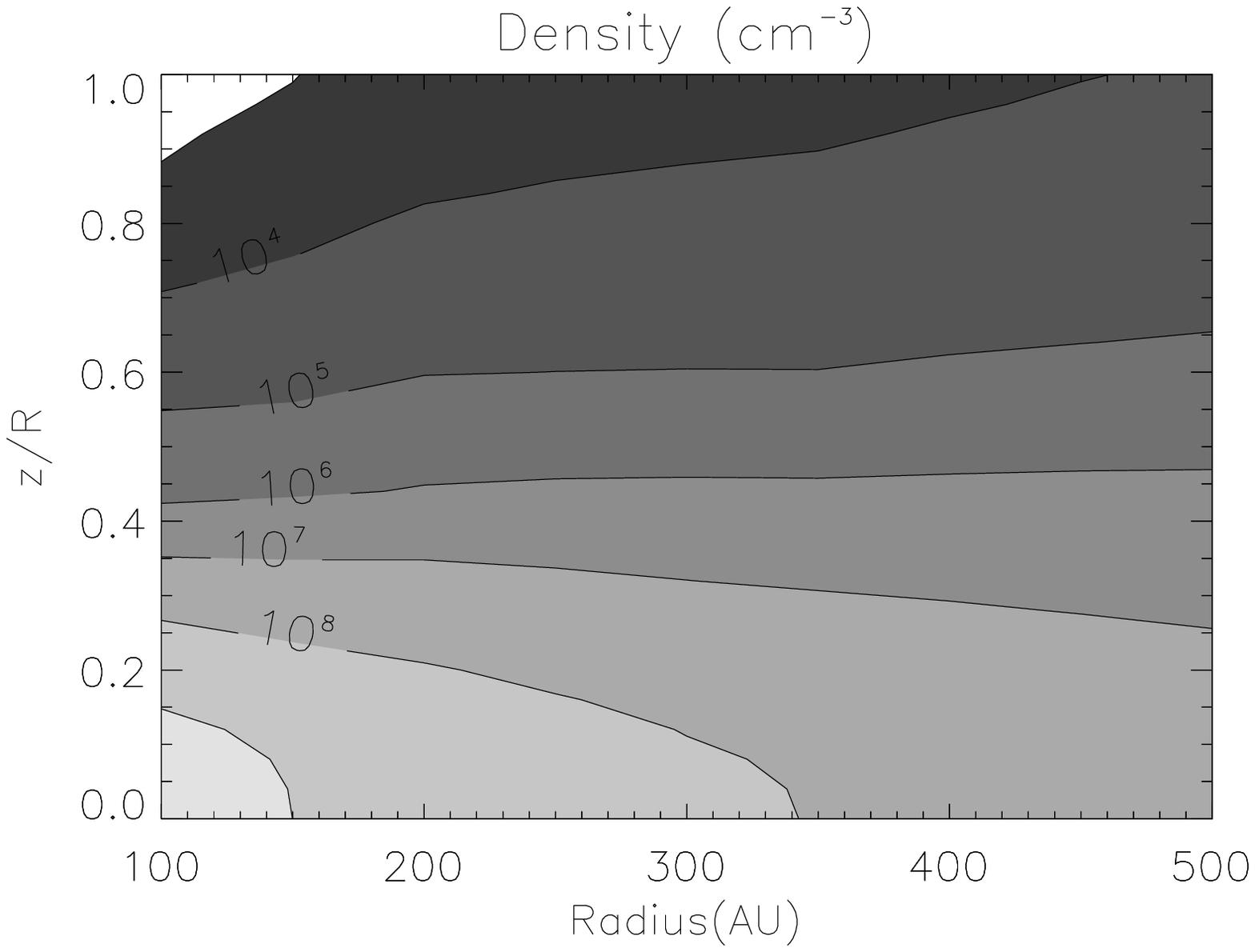}{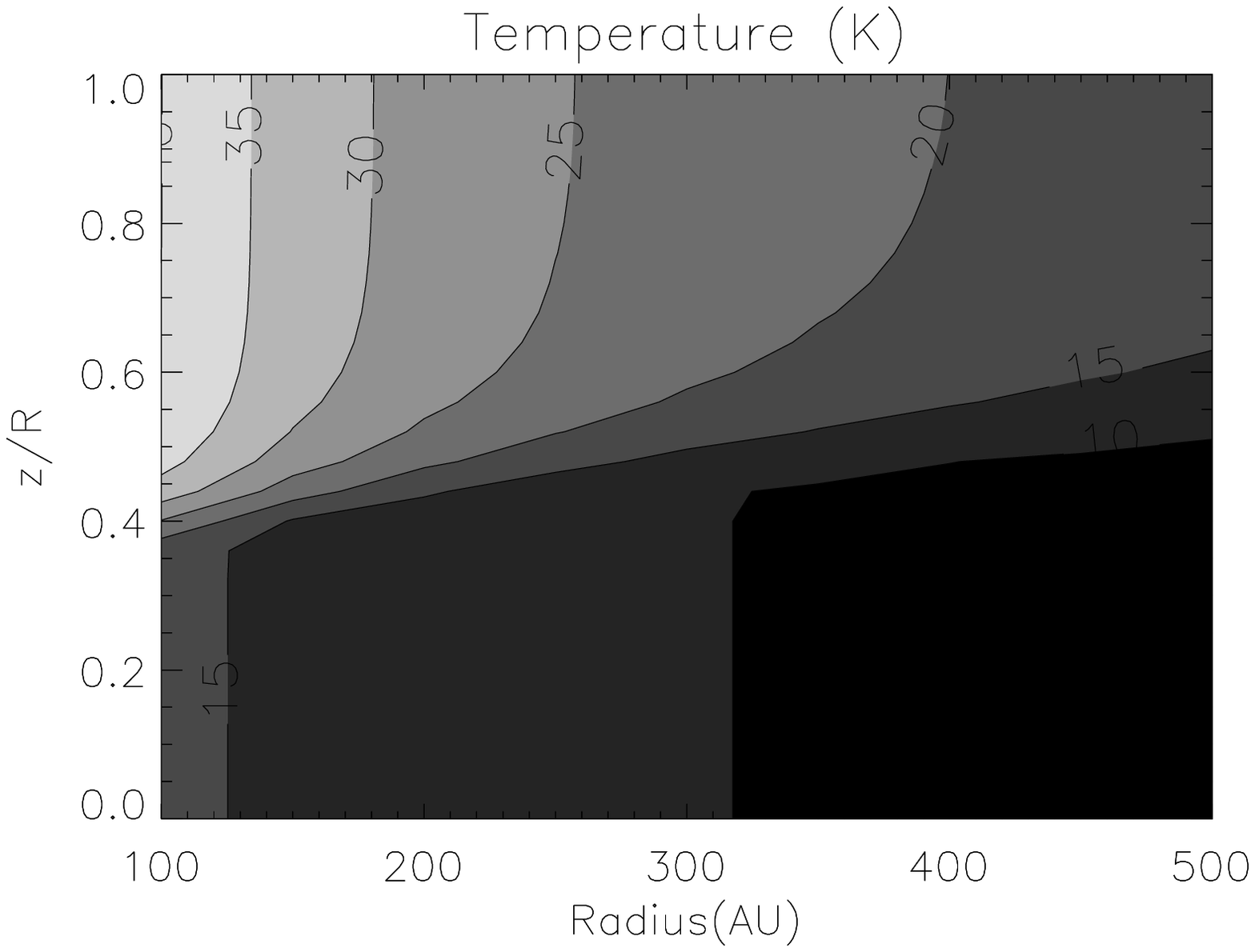}
\caption{\label{fig:disk}The physical structure of a disk calculated
using a hydrodynamical model of a geometrically thin disk heated
both viscously and by irradiation from the central star.  The 
equations are solved semi-analytically.  The disk is assumed to
be in steady state.  The mass accretion rate is 10$^{-8}$
M$_\odot$ yr$^{-1}$, $\alpha$ = 0.01, $\Sigma_0$ = 1000 g cm$^{-2}$ at 1 AU
\citep{bryden06}.  {\it Left Panel}: Number density profile.
Contours run from 10$^3$ to 10$^9$ cm$^{-3}$. {\it Right panel}: Grain
temperature (assumed to be equal to the gas temperature).  Contours
are at 10, 15, 20, 30, 40, 50, 60 and 70 K.}
\end{figure}

The calculated density and temperature structure from this model are shown
in Figure~\ref{fig:disk}.   We assume that the gas temperature 
is the same as the grain temperature.
In reality, this probably underestimates the gas temperature
in the surface layers \citep{kd04}.  
As has been seen in previous models, the temperature
structure of the disk can critically affect the molecular distributions
in the disk by controlling the regions where thermal desorption occurs.
The grain temperature distribution in our model is consistent with that
calculated by \citet{dalessio01} although somewhat colder
than in previous models by \citet{dalessio99}.  
The reduction in temperature arises from the different assumptions
about the grain size distribution.  Both \citet{dalessio01} and
\citet{bryden06} assume that some grain coagulation has
taken place so that the grain size distribution is closer to
a Pollack distribution than the interstellar grain distribution
assumed to \citet{dalessio99}.  These colder disk models have
been found to be in better agreement with the observed SEDs of disks.

%The difference
%is due to the assumptions made about the grain size distribution.
%\citet{dalessio01} and \citet{bryden06} both assume that some
%grain coagulation has taken place, whereas \citet{dalessio99}
%uses an interstellar dust distribution.  With the larger grains 
%the resulting disks are thinner and have
%colder dust temperatures in the surface layers.  These colder
%disk models have been found to produce better agreement with the 
%observed spectral energy distributions (SEDs) of T Tauri stars.  However the lower grain temperature
%has important implications for the chemistry, with thermal desorption being
%less efficient, resulting in low gas phase column densities
%of some species.  Because of this low temperature we have, in previous models,
%invoked a non-thermal desorption process e.g.\ photodesorption in 
%\citet{wl00}, to maintain observable levels of molecules
%in the gas.  Here we exclude desorption processes other than
%thermal desorption and cosmic ray heating because
%we wish to evaluate only the effects of diffusion.
%(Cosmic ray heating only affects the abundances of the most
%volatile molecules such as N$_2$ and CO).
%Other disk chemical models, e.g.\ \citet{aikawa02} use the older d'Alessio
%et al.\ models with warmer disks 
%and therefore do not have the same issues of low temperatures
%and resulting low column densities.

\section{Chemistry}
\subsection{Reaction set and input abundances}

The chemical model is a subset of the UMIST RATE95 database
\citep{rate95}. 
We have followed the chemistry of 79 gas phase and  31 ice
species (containing H, He, C, N, O and Si), linked by 1480 reactions.  
The full species set is listed in Appendix~\ref{ap:spec}.
The chemistry includes thermal desorption, 
ionization (from the interstellar radiation field
and the star, and from the decay of radioactive isotopes), and
gas--grain interactions, 
including surface reactions and thermal desorption.  These processes
are discussed in more detail below.

To take into account the fact that material
incorporated into the disk has already undergone processing
in the parent molecular cloud we take the input abundances
to the disk model from the output of a molecular cloud model
that has been allowed to run for 1 Myrs. The input
abundances for the cloud model are given in Table~\ref{tab:atoms}.
We assume all elements are in their atomic form (or ionic in the case
of carbon), with the exception of hydrogen which is 99\% molecular.
The cloud model is run with a density of 2 $\times$ 10$^4$ cm$^{-3}$,
at a temperature of 10 K and with a visual extinction of 10 magnitudes.
The species and reaction sets used are the same as for the the disk model.
Ionization is provided by cosmic rays, photons and cosmic ray
induced photons, and freezeout and desorption (thermal and
cosmic ray heating of grains) is included.  This results
in a mixture of gas and ice at 1 Myrs.  The abundances
of the major species at this time are listed in Table~\ref{tab:init}
and these are used as inputs to the disk models discussed below.

\begin{deluxetable}{ll}
\tablecolumns{2}
\tablewidth{0pt}
\tablecaption{\label{tab:atoms}The input abundances
for the molecular cloud model.  Since a protoplanetary disk
forms from material that has been processed in a molecular
cloud, we use the outputs from the cloud model
at 1 Myrs to provide the input abundances for the disk model.
These are listed in Table~\ref{tab:init}.}

\tablehead{
\colhead{Element} & \colhead{Fractional abundance} 
}
\startdata
H$_2$ & 0.99 \\
H & 0.01 \\
He & 0.14\\
O & 1.76 $\times$ 10$^{-4}$ \\
C$^+$ & 7.3 $\times$ 10$^{-5}$ \\
N & 2.14 $\times$ 10$^{-4}$ \\
Si & 2.0 $\times$ 10$^{-8}$
\enddata
\end{deluxetable}

\begin{deluxetable}{lll}
\tablecolumns{3}
\tablewidth{0pt}
\tablecaption{\label{tab:init}Initial abundances as calculated in
a molecular cloud model.  a (b) indicates {a~$\times$~10$^b$}.  }
\tablehead{
\colhead{Species} & \multicolumn{2}{c}{Fractional abundance} \\
& \colhead{Gas} & \colhead{Grain}
}
\startdata
CO         & 6.2 (-6) & 6.0 (-5) \\
CO$_2$     & 1.3 (-8) & 2.4 (-7)\\
H$_2$CO    & 3.8 (-8) & 1.3 (-9) \\
O$_2$      & 6.0 (-9) & 1.3 (-5) \\
H$_2$O     & 1.5 (-7) & 7.5 (-5) \\
CH$_4$     & 1.6 (-7) & 5.3 (-7)\\
C$_2$H     & 1.1 (-8) & \\
C$_2$H$_2$ & 7.9 (-8) & 1.6 (-6) \\
CN         & 3.0 (-8) \\
HNC        & 2.7 (-8) & 7.2 (-7) \\
HCN        & 4.2 (-8) & 6.0 (-7) \\
HC$_3$N    & 6.0 (-8) & 3.7 (-7)\\
NH$_3$     & 1.7 (-9) & 5.3 (-6) \\
N$_2$      & 4.8 (-6) & 5.5 (-8) \\
CH$_3$CN   & 1.7 (-9) & 8.2 (-9) \\
CH$_3$OH   & 1.6 (-10) \\
N          & 2.8 (-7) \\
O          & 2.2 (-6) \\
HCO$^+$    & 7.6 (-10) &\\
N$_2$H$^+$ & 5.0 (-10) &  \\
\enddata
\end{deluxetable}

\subsection{Ionization processes}

Several processes contribute to ionization in the disk.  As
in molecular clouds, cosmic rays can cause ionization if the
surface density is low enough for them to penetrate (cosmic
rays can penetrate densities up to 150 gcm$^{-2}$ \citep{un81}).  This is
always the case in the region of disk considered here.  Indirectly,
cosmic rays are responsible for a UV photon field which 
can cause photodissociation \cite{pt83}.  Cosmic
rays ionize H$_2$, generating secondary electrons with an energy
of $\sim$ 30 eV.  These electrons can excite electronic states
of H$_2$ which then decay producing a UV flux.

UV photons from both the star and the interstellar radiation field
can cause ionization in the disk. We estimate the UV field
at any given position in the disk by assuming that
interstellar photons hit the disk vertically and that stellar
photons travel horizontally.  The strength of the stellar UV field
for a T Tauri star has been estimated as 10$^4$ times the interstellar radiation field at 100
AU by \citet{hg96} and more recently as a few hundred
times the interstellar radiation field at 100 AU by \citet{bergin03}.
Here we have chosen to use a value of 500 times the interstellar
radiation field
following Bergin et al.  In common with \citet{aikawa02}, we
have assumed that the stellar radiation field does not dissociate CO
and \chem{H_2}, although these molecules are dissociated by 
the interstellar radiation field.  We use the approach of
\citet{lee96} to describe the self--shielding of these molecules.

The decay of radioactive isotopes can be a source of ionization 
in the disk \citep{un81}.  $^{26}$Al can decay to form excited $^{26}$Mg
which in turn decays by positron decay or by electron capture.  
We include the
ionization due to these decays using the rate found by
\citet{un81} i.e.\ $\zeta_R$ =  6.1 $\times$ 10$^{-18}$
s$^{-1}$.

T Tauri stars can have strong X-ray fields.  
\citet{ig99}
modeled the effects of X-ray ionization in the disk
and found that it is effective for $R$ $<$ 10 AU.  It is most
effective in the surface layers because the attenuation length of
X-rays is very small.  \citet{ah01} included X-rays but
found that they did not significantly alter the results for the outer
disk.  Since we are considering only $R$ $>$ 100 AU we have chosen to
ignore the effects of X-rays.

\subsection{Gas--grain interactions and grain surface chemistry}

We assume that all species can freezeout onto grains.  
The exception is He which has such a low binding energy
(100 K; \citet{th82})
that it is easily desorbed even at very low temperatures.
We therefore assume that He is not affected
by collision with grains and any He$^+$ that collides with
a grain is neutralized
and returned to the gas.
For all other species the
freezeout rate is given by
\begin{equation}
k_f = S_x <\pi a^2 n_g> C_i v_x n_x
\end{equation}
where $S_x$ is the sticking coefficient (assumed to be 0.3 for
all species), $a$ is the
grain radius, $n_g$ is the number density of the grains,
$v_x$ is the gas phase velocity of species $x$ and $n_x$ 
is the number density of $x$.  
We assume $< \pi a^2 n_g >$ = 2.1 $\times$ 10$^{-21}$ $n_H$ cm$^{-1}$, and
$n_g$ = 10$^{-12}$ $n_H$ cm$^{-3}$.
$C_i$ is a factor that
takes into account the effect of charge on the
freezeout rate.
\cite{un80} showed that the majority of grains in 
dense regions are negatively charged.  Hence
the accretion rate of a positively charged ion 
is enhanced compared to that of neutrals and $C_i$ is given by
\begin{eqnarray}
C_i  & = & 1 \mbox{~~~ for neutral species} \nonumber\\
     & = & 1 + e/(akT) \mbox{~~~ for single charged
positive ions}
\end{eqnarray}
where $e$ is the electron charge and $k$ is the
Boltzmann constant.  (We do not explicitly track the
grain charge, but instead following \cite{un80} assume that
the grains on average have a negative charge).
Ions that hit a grain are neutralized in the same way
that they would be if they reacted with an electron in the gas.

Reactions between grain--surface species are included.
We assume that only atoms are able to freely move across
the grain surface, larger species are prevented from
moving because of their higher binding energies.  Therefore
all grain surface reactions involve at least one atom. 

The rate equation method is still the most tractable means
of modeling the combined gas and grain chemistry.
However, it is know to be flawed because  it does not
take into account the discrete nature of grains, with
potentially significant effects on the calculated abundances.
It has been shown that the results for a rate equation
calculation of the ice composition can differ markedly 
from those of a Monte Carlo model \citep{charnley97,tc97,caselli98}.
One way around this is to reduce the scan rate of hydrogen atoms
across the surface of a grain, thus reducing their reaction rate,
and bringing the results of the two methods into better agreement.
Recent experimental work by \cite{katz99} has suggested
that the scan rate of hydrogen atoms is much lower than was generally
assumed.  \cite{rh00} found that a rate equation model using
the scan rates of \cite{katz99} gives better agreement with the
Monte Carlo models, than do models which utilize the fast scan rate.
Here we have chosen to follow \cite{rh00} in using the binding energy
data of \cite{katz99} to calculate the grain surface reaction rates.
The binding energy data used is given in Table~\ref{tab:be}.

The grain surface reaction set is taken from \cite{hh93}.
The rates are assumed to be temperature dependent and are calculated
as follows.  The scan rate of an atom can be calculated from
\begin{equation}
t^{-1} = \nu_0 exp(-E_b/kT_{gr})
\end{equation}
where $\nu_0$ is the frequency of oscillation between absorbate
and surface, given by
\begin{equation}
\nu_0 = \sqrt (2 n_s E_D/ \pi^2 m)
\end{equation}
where $n_s$ is the surface density of sites ($\sim$ 1.5 $\times$ 10$^{15}$ cm$^{-2}$),
$E_D$ is the binding energy of the atom, $m$ its mass, $E_b$ = 0.3 $E_D$ 
and $T_{gr}$ is the grain temperature.  

Molecules in the ice mantle are returned to the gas phase
by thermal desorption and by desorption due to cosmic ray
heating of grains.
The rates for thermal desorption are calculated using the
binding energies listed in Table~\ref{tab:be}.
The rates for cosmic ray heating are calculated using the
method of \citet{hh93}.
We find that for the majority of species cosmic ray heating
is not an important process. The exceptions to this are those
molecules which are relatively weakly bound e.g. N$_2$.

\begin{deluxetable}{lrc}
\tablecolumns{4}
\tablewidth{0pt}
\tablecaption{\label{tab:be}The binding energies ($E_b$) used
to determine the thermal desorption rates of the abundant
mantle species. References are (1) \citet{sa93}, (2) \citet{sa90},
(3) \citet{sa88}, (4) \citet{ta87}, (5) \citet{rh00}}
\tablehead{
\colhead{Species} & \colhead{Binding} & \colhead{Reference} \\
& \colhead{Energy (K)} & 
}
\startdata
H        & 373  & 5 \\
C        & 800  & 4 \\
N        & 800  & 4 \\
O        & 800  & 4 \\
CO       & 1210 & 4 \\  
CO$_2$   & 2860 & 2 \\
H$_2$O   & 4820 & 3 \\
NH$_3$   & 3080 & 2 \\
H$_2$CO  & 1760 & 4 \\
CH$_3$OH & 4240 & 1 \\
H$_2$    & 315  & 5 \\
N$_2$    & 710  & 4 \\
O$_2$    & 1210 & 4 \\
\enddata
\end{deluxetable}

\section{Results}

\begin{figure}
\epsscale{0.9}
\plotone{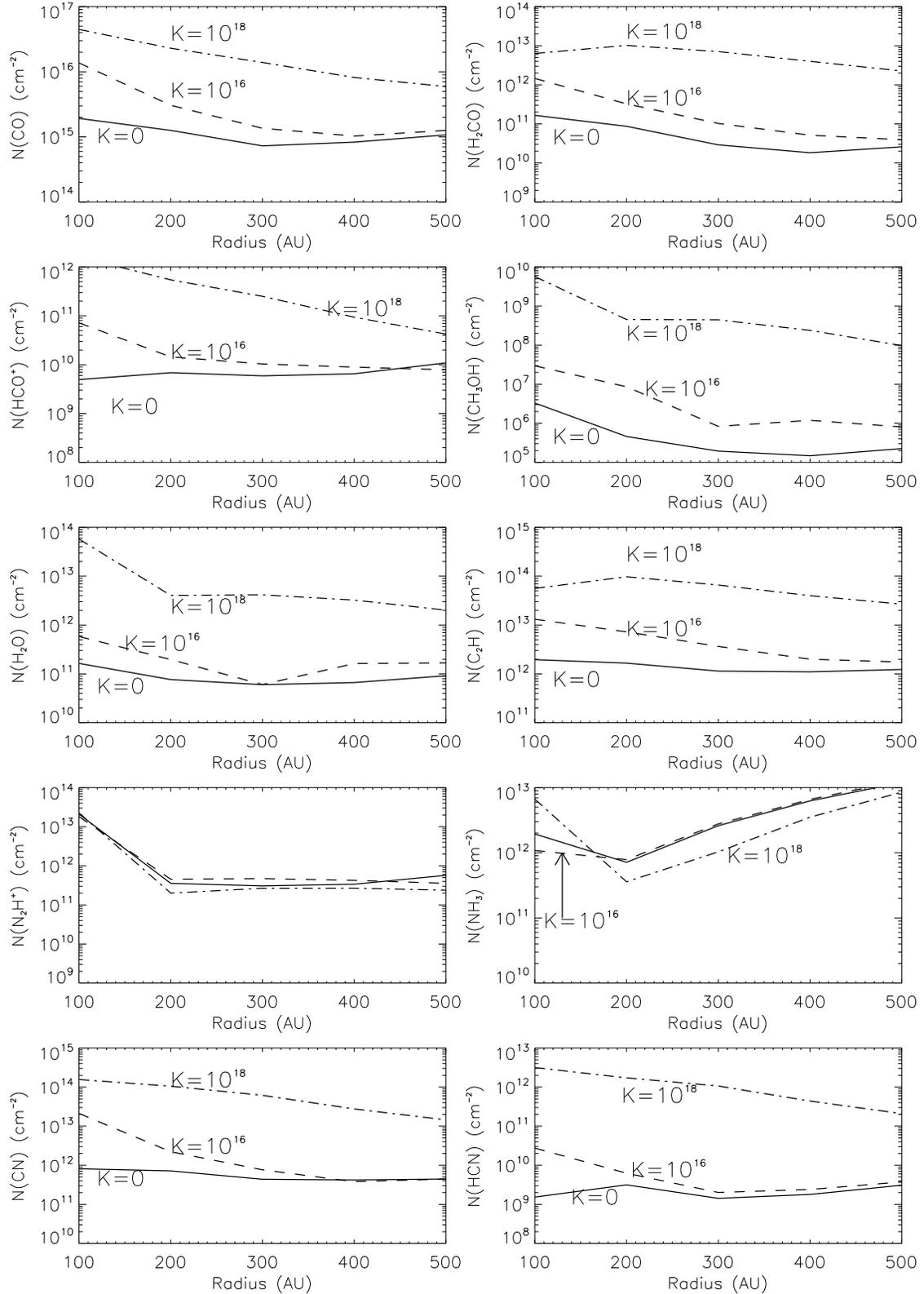}
\caption{\label{fig:cd}Radial distribution of the 
column densities through the disk for several
molecules at a model time of 1 million years. The solid lines indicate $K$ = 0 cm$^{2}$s$^{-1}$,
dashed lines are for $K$ = 10$^{16}$ cm$^2$s$^{-1}$ and dash-dotted lines
are for $K$ = 10$^{18}$ cm$^2$s$^{-1}$.  These figures show that
diffusion increases the column densities of most species.  The exceptions
are \chem{NH_3} and \chem{N_2H^+}, which are most abundant in the midplane rather
than in the molecular layer.}
\end{figure}

The effects of diffusion on the calculated fractional abundances and
column densities depends on the position in the disk and
on the molecule under consideration.  
Figure~\ref{fig:cd} shows the calculated column densities
at 1 Myrs for several important molecules and illustrates how
some molecules are greatly affected by the inclusion of 
diffusion, whereas others, notably \chem{N_2H^+} and \chem{NH_3}
are relatively unaffected.  The effect increases with increasing
$K$, and the column densities for $K$ = 10$^{16}$ cm$^2$s$^{-1}$
are very similar to $K$ = 0 for $R$ $>$ 200 AU.  The effect is
increased at smaller $R$ because of the higher disk temperature
which increases the abundance of molecules in the gas as a result
of thermal desorption.  For $K$ = 10$^{18}$ cm$^{2}$s$^{-1}$ 
a similar radial effect is seen.  

\begin{figure}
\epsscale{0.9}
\plotone{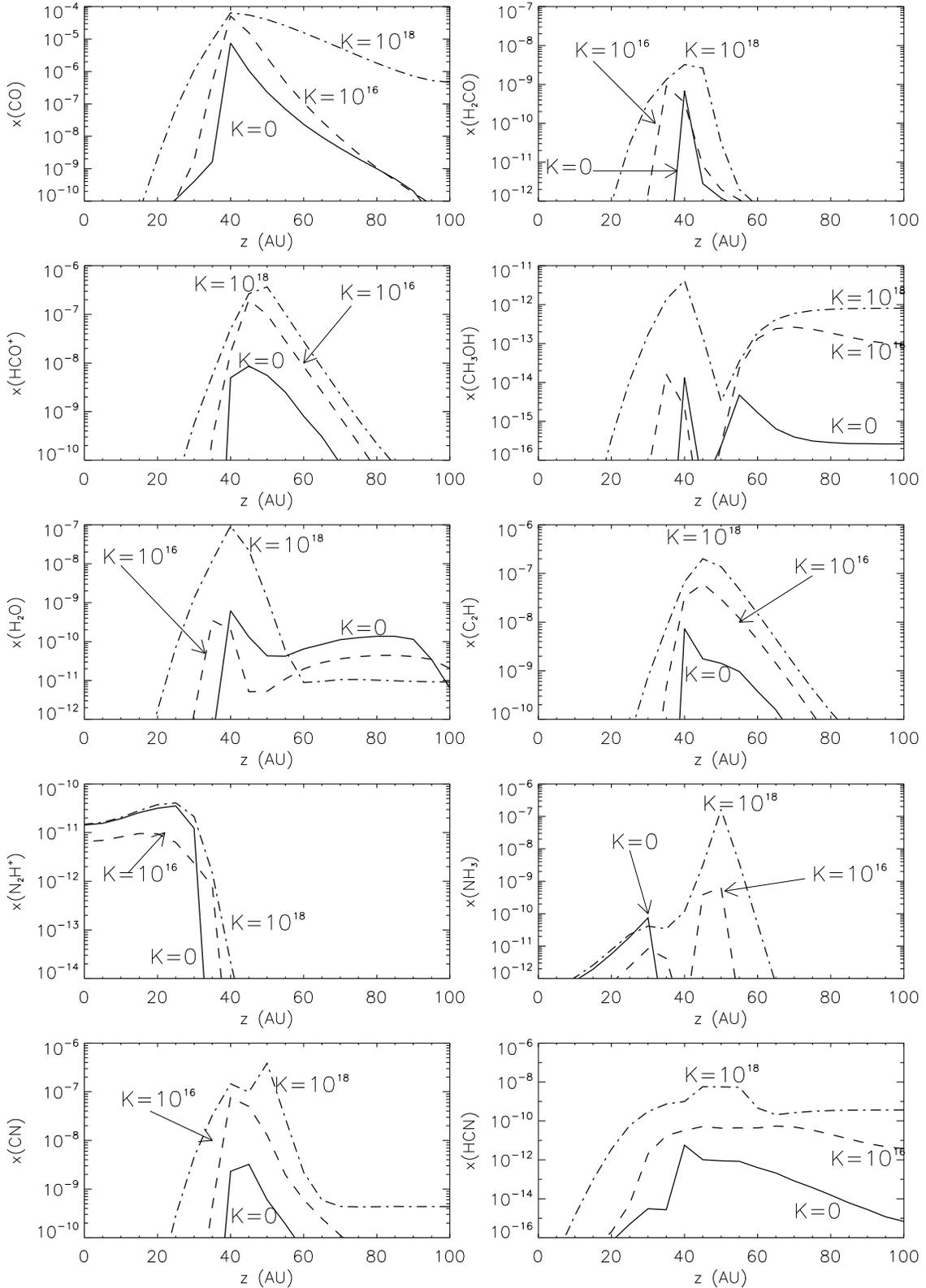}
\caption{\label{fig:frac100} Fractional abundances 
with respect to H$_2$ as a function of 
height $z$ above the midplane for $R$ = 100 AU and a model time of 1 million years.
The solid lines indicate $K$ = 0 cm$^{2}$s$^{-1}$,
dashed lines are for $K$ = 10$^{16}$ cm$^2$s$^{-1}$ and dash-dotted lines
are for $K$ = 10$^{18}$ cm$^2$s$^{-1}$.  Diffusion increases the peak abundance
and the vertical extent of many molecules.  At this radius, \chem{N_2H^+} 
traces the midplane, but \chem{NH_3} peaks in the molecular layer.}
\end{figure}

\begin{figure}
\epsscale{0.9}
\plotone{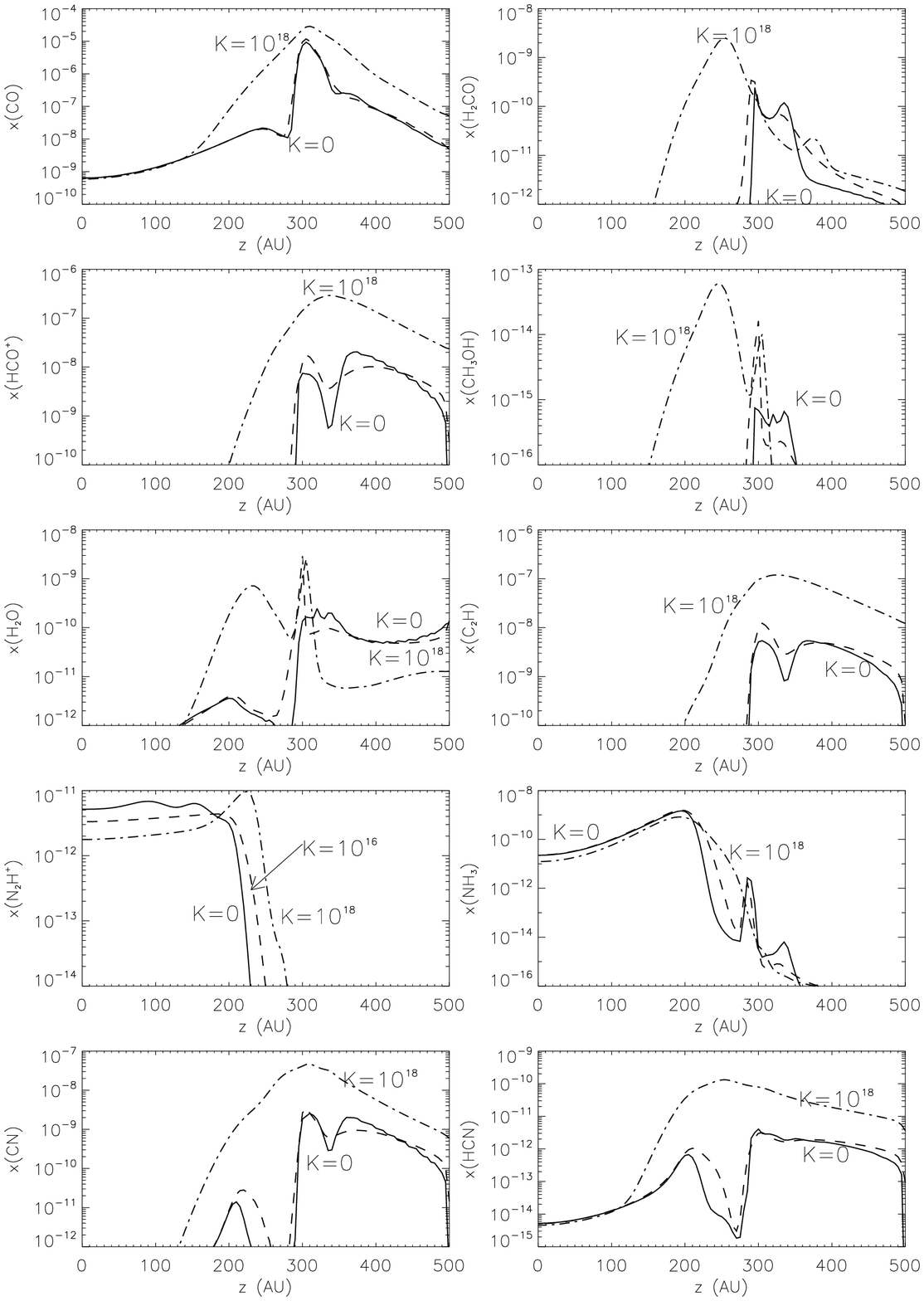}
\caption{\label{fig:frac500} Fractional abundances as a function of 
height $z$ above
the midplane for $R$ = 500 AU and a model time of 1 million years.
The solid lines indicate $K$ = 0 cm$^{2}$s$^{-1}$,
dashed lines are for $K$ = 10$^{16}$ cm$^2$s$^{-1}$ and dash-dotted lines
are for $K$ = 10$^{18}$ cm$^2$s$^{-1}$.  At this
radius $K$ = 10$^{18}$ cm$^2$s$^{-1}$ is required to affect the
abundances and to increase their peak values and the vertical extent
of most molecules.  \chem{NH_3} and \chem{N_2H^+} both trace the midplane,
whereas other molecules trace the molecular layer.}
\end{figure}

In general, diffusion smooths out
abundance variations with $z$ (as expected) and increasing the depth of the molecular
layer (Figures~\ref{fig:frac100} and \ref{fig:frac500}).  The chemical
effects induced by diffusion arise both from the transport
of icy grains from the cold midplane into a warmer region, where
the ices can be desorbed, and by the movement of atoms and ions
from the highly irradiated surface layer into more shielded regions
where molecules can form and survive.  Molecules in the
diffusion models are formed by the same reactions as in the
no diffusion case, but the abundances can be increased because
of the increase in potential reactants.

Most molecules exist in a layer between the cold midplane and
the UV irradiated surface region (Figure~\ref{fig:disk_chem}).  
Exceptions to this are \chem{N_2H^+} and \chem{NH_3} which
have peak abundances in the midplane.  The upper bound of
the molecular layer is determined by photodissociation effects, and the lower
one by the point where freezeout dominates the formation 
processes.
For  molecules in this layer,
the transport of atoms and their ions from the surface
is important in driving the chemistry. Diffusion brings
atomic and ionized carbon, nitrogen and oxygen into the molecular
layer, increasing the formation rate of molecules such as CO, \chem{H_2O}
and \chem{CH_4}.  These form in the gas phase by the same 
ion-molecule reaction
routes as are found in molecular clouds.  For CN and HCN
however, neutral--neutral reactions are important:
\begin{equation}
\hbox{CH} + \hbox{N} \longrightarrow \hbox{CN} + \hbox{H}
\end{equation}
\begin{equation}
\hbox{CH$_2$} + \hbox{N} \longrightarrow \hbox{HCN} + \hbox{H}
\end{equation}
The usual gas phase reaction formation via HCNH$^+$ is only
important in the $K$ = 0 cm$^2$s$^{-1}$ case where it is comparably
efficient to the neutral--neutral processes.

We consider CO to illustrate how diffusion alters the chemistry.
$N$(CO) increases with $K$ for all radii.
The same reactions form and destroy this molecule in all models,
CO is formed by thermal desorption and destroyed
by freezeout in the midplane region, and by photodissociation in 
the outer layers.  The changes in column density
are due to the effects of diffusion which broadens the depth of the 
molecular layer and increases the peak fractional abundance 
(Figure~\ref{fig:frac100}).  The transport of icy grains towards the surface
ensures that there is a continual source of CO in the diffusion
models, leading to higher CO fractional abundances and a broader
high abundance region.  At 100 AU the peak abundance of CO occurs
at $z$ = 40 AU.  For $z$ $>$ 40 AU thermal desorption ensures that
the grains are free of CO ice.
In this region there is an additional contribution 
to the CO from gas phase chemistry for $K$ = 10$^{18}$ cm$^2$s$^{-1}$
from the dissociative recombination of HCO$^+$.  Similar effects
are seen at other radii (Figure~\ref{fig:frac500}).

Water is another molecule whose column density is increased by
diffusion.  In this case the oxygen chemistry is driven
by the transport of grains into the warm surface region
where O$_2$ ices can desorb.  O$_2$ can form efficiently on 
the grains near the surface in regions where oxygen atoms are
plentiful but the hydrogen is mainly tied up in H$_2$. 
Photodissociation of O$_2$ can then produce
oxygen atoms which are transported in turn into the molecular layer.
Another source of oxygen atoms is the photodissociation of CO.
The direct desorption of water ice is not a good source of gas phase
water as the temperature is too low for thermal desorption to occur 
at the disk radii considered here.  

\chem{N_2H^+} and \chem{NH_3} differ from many other species
in that they  have high midplane abundances.  They also show
little variation in column density with diffusion.  Both molecules
are derived from N$_2$ (\chem{N_2H^+} forms
from the reaction of \chem{N_2} with \chem{H_3^+}) whereas
\chem{NH_3} is formed by a chain of reactions starting from the breakup
of N$_2$ into N$^+$ and N atoms by reaction with He$^+$.  
N$_2$ is very volatile and can be desorbed
even in the midplane (either by thermal desorption or by cosmic
ray heating).  It therefore cycles back and forth between the 
gas and solid phases.  Diffusion means that both gas and solid N$_2$ 
are transported, spreading out the molecules so that the peak
abundance falls, and the vertical extent of the N$_2$ layer 
is increased slightly.  Hence we see some small changes
in the vertical fractional abundance distribution, but
relatively little change in the column densities.

The ionization fraction $x(e)$ is not significantly affected
by diffusion.  At 100 AU we find $x(e)$ = 2.5 $\times$ 10$^{-11}$ for
models with both $K$ = 0 and $K$ = 10$^{18}$ cm$^2$s$^{-1}$.  
At 300 AU the ionization fraction is 1.45 $\times$ 10$^{-10}$ for $K$ = 0 and 
2.0 $\times$ 10$^{-10}$ for $K$ = 10$^{18}$.  At 500 AU
a slight decrease in $x(e)$ from 7.9 $\times$ 10$^{-10}$ to 6.3 $\times$ 10$^{-10}$
is seen when diffusion is included.  Recent observations
by \citet{cec04} have estimated $x(e)$ in the midplane of a disk
and we compare our results to these observations below.

In summary, most molecules show an increase in column density
with increasing diffusion.  Those with peak abundances that 
lie near the midplane (e.g.\ molecules such as NH$_3$ and
\chem{N_2H^+} which can remain in the gas even after 
other molecules are heavily depleted), are least affected
by diffusion.  Those molecules that peak in the molecular
layer (Figure~\ref{fig:disk_chem}) are affected by the
transport of ices into the warmer layers where they can
be desorbed and photodissociated and returned as reactive atoms and ions
to the molecular layer, where they can drive the chemistry.
Our models indicate that \chem{N_2H^+} and \chem{NH_3} 
are important tracers of the midplane conditions and dynamics
in the outer disk.  As the radius decreases NH$_3$ moves into
the molecular layer, but \chem{N_2H^+} remains in the midplane (see
Figures~\ref{fig:frac100} and \ref{fig:frac500}).

\subsection{Comparison to observations}

We present a comparison of our results with available observations of
the outer parts of the disks around the T Tauri stars, DM Tau, LkCa 15
and TW Hya.  DM Tau is located 140 pc away in Taurus; with an age of
$\sim$ 5 Myrs, it is one of the oldest T Tauri stars.  It has a
stellar mass of 0.55 M$_\odot$ \citep{simon00} and a temperature of
3720 K \citep{gd98}.  The disk itself has a gas mass of 0.03 M$_\odot$
\citep{gd98}, a radius of $\sim$ 800 AU \citep{simon00}, and a mass
accretion rate is 10$^{-8}$ M$_\odot$ yr$^{-1}$\citep{hartmann98}.
LkCa 15 is also located in Taurus.  The star has a mass of 1 M$_\odot$
\citep{simon00} and a temperature of 4365 K \citep{muz00}.  The disk
radius is slightly smaller than that of DM Tau at 650 AU
\citep{simon00} and it has a mass accretion rate of 10$^{-9}$
M$_\odot$ yr$^{-1}$\citep{hartmann98}.  The age has been variously
estimated at 3--5 Myrs \citep{simon00} to 12 Myrs \citep{thi01}.  TW
Hya is a classical T Tauri star with a nearly face on disk at $\sim$
56 pc \citep{webb99}.  Its age is $\sim$ 15 Myrs and its mass
accretion rate is $\sim$ 10$^{-8}$ \citep{kastner02}.

The column densities derived from observations are very sensitive to
assumptions made about the source.  Single dish observations cannot
resolve the disks and assumptions about the density and temperature
structure of the disks must be made, in order to determine abundances
and column densities.  For DM Tau \citet{dutrey97} determined the gas
density distribution of a geometrically thin disk in hydrostatic
equilibrium and used this to derive the average fractional abundances
with respect to H$_2$, assuming that the fractional abundances were
the same everywhere in the disk.  \citet{aikawa02} used the model and
data from \citet{dutrey97} and integrated vertically to determine the
column densities quoted in Table~\ref{tab:col}.  For LkCa 15 both
single dish and interferometric data is available.  \citet{qi01} used
the Owens Valley Radio Interferometer (OVRO) array to observe this source and
derived beam average column densities.  The resolution of the array
($\sim$ 300 AU at the distance of LkCa 15) is such that the source is
just resolved.  \citet{thi04} present single dish observations of LkCa
15.  They derive the column densities assuming that the radius of the
disk is 450 AU.  The disk radius can have a large effect on the
calculated column densities with a change from $R_1$ to $R_2$ scaling
the column densities by $(R_1/R_2)^2$.  The single dish column
densities are significantly higher than those from the interferometer
data due to the differences in the assumptions used to determine these
numbers.

%Given the uncertainties in both the models and the observations the
%agreement is remarkably good for DM Tau.  We find good agreement for
%all the available data for $K$ = 10$^{18}$ cm$^{2}$s$^{-1}$.  (see
%Table~\ref{tab:col}), where good agreement is defined as being within
%a factor of 5 of the observed column density).  Diffusion ($K$ =
%10$^{18}$ cm$^2$s$^{-1}$) increases $N$(CO) to 1.3 $\times$ 10$^{16}$
%cm$^{-2}$, $N$(\chem{N_2H^+}) is consistent with the observed upper
%limit.  For Lk Ca 15 We are able to match the single dish
%observations, but have less success with the interferometric data.  In
%the case of TW Hya, our diffusion model does not produce sufficient
%HCN or HCO$^+$ and overproduces H$_2$CO.  In the case of the
%interferometer observations of LkCa 15 we are unable to account for
%the column densities of HCN, CH$_3$OH and HCO$^+$, but we find good
%agreement for other molecules.

Given the uncertainties in both the models and the derivation
of column densities from the observations, our model results for 
$K$ = 10$^{18}$ cm$^{2}$s$^{-1}$ are in remarkably good agreement with the observations of
DM Tau.  We find good agreement (defined as the column densities from the
model and observations agreeing to within a factor of 5) for all observed
molecules in this source.  For Lk Ca 15 we are able to match the single
dish observations, but have less success with the interferometric data, 
which tend to result in much larger column densities.  Consequently our
$K$ = 10$^{18}$ model is unable to account for the column densities of
HCN, CH$_3$OH and HCO$^+$, although we find good agreement for the other
molecules.  For TW Hya, our $K$ = 10$^{18}$ model does not produce
sufficient HCN or HCO$^+$, and overproduces H$_2$CO.  

%Our calculated $N$(CH$_3$OH) is extremely low.  
\chem{CH_3OH} has been
detected in LkCa 15 by \citet{qi01} using OVRO 
with a column density of 7.3 $\times$ 10$^{14}$ --
1.8 $\times$ 10$^{15}$ cm$^{-2}$.  \citet{thi04} found an upper limit for
$N$(\chem{CH_3OH}) in the same source of $<$ 9.4 $\times$ 10$^{14}$ cm$^{-2}$.  
H$2$CO has also been observed in LkCa15 and the column density is
estimated to be
1.0 $\times$ 10$^{14}$ cm$^{-2}$ by \cite{thi04} or 7.2 $\times$ 10$^{12}$ --
1.9 $\times$ 10$^{13}$ cm$^{-2}$ by \cite{aikawa03}.  Our model 
value of $N$(H$_2$CO) = 7.1 $\times$ 10$^{12}$ 
at $R$ = 300 AU for $K$ = 10$^{18}$ cm$^2$s$^{-1}$ is
in agreement with the observations in this source, 
but we are unable to account for the observed $N$(CH$_3$OH).
Both \chem{H_2CO} and
\chem{CH_3OH} are thought to form on the surfaces of dust grains by
the hydrogenation of CO, although the exact mechanism and the reaction
rates are still a matter of debate \citep{wk02, hiraoka02}.
Due to the lack of consensus on the rates we have not included 
the formation of CH$_3$OH on grains, although we do have 
formation of H$_2$CO from hydrogenation of CO with an activation barrier
of 1000 K.  This results in the formation of a small amount of H$_2$CO
on the grains.  Diffusion increases the column density of
both molecules, with a slightly greater effect on \chem{CH_3OH}.
The inclusion of grain formation reactions for CH$_3$OH would 
increase its column density somewhat but its high binding energy
($E_b$(\chem{CH_3OH}) = 4240 K: \cite{sa93}) means that it is not
easily returned to the gas unless there is a non--thermal desorption
process acting. The lower binding energy of H$_2$CO compared to CH$_3$OH
also means that it is returned relatively easily to the gas in the warmer
regions of the disk.  

The ionization fraction, $x(e)$, is an important parameter in disk
modeling, possibly controlling the turbulence in a disk.  
\citet{cec04} observed \chem{H_2D^+} in several T Tauri stars
and from this estimated the value of $x(e)$ in the midplane to be 7
$\times$ 10$^{-10}$ in DM Tau, 4 $\times$ 10$^{-10}$ in TW Hya and $<$
2 $\times$ 10$^{-9}$ in LkCa 15.  (\chem{H_3^+} and its
deuterated isotopes are likely to major carriers of charge
in the midplane where the high densities and low temperatures mean
that other molecular ions e.g.\ HCO$^+$ are heavily depleted). 
Our results find a value $x(e)$ $\sim$ a few $\times$ 10$^{-10}$ 
in the outer disk midplane, with the exact value depending on the radius.
This is consistent with the observed values for DM Tau and TW Hya
but a little lower than that estimated for LkCa15.
We find that the inclusion of diffusion does not affect $x(e)$ and
that the major charge carriers in the midplane are the ions of
atomic and molecular hydrogen (H$^+$ and \chem{H_3^+}).

We find that our static ($K$ = 0 cm$^2$s$^{-1}$) models cannot keep sufficient
molecules in the gas phase to account for the observed column
densities without the inclusion of additional diffusion processes
such as photodesorption.
This result is a consequence of the relatively cool disk model we use
for the density and temperature distributions (\citet{aikawa02} use a
warmer disk model which permits the efficient thermal desorption of
volatile molecules, and find much higher column densities as a
result).  The cooler models predict spectral energy distributions that
are in better agreement with the observed T Tauri disks, but cause
problems for the understanding of the chemistry.  For example, 
observations of DM
Tau find $N$(CO) = 5.7 $\times$ 10$^{16}$ cm$^{-2}$ and $N$(CO) is
even higher in LkCa 15 where $N$(CO) =  1.6 $\times$ 10$^{18}$ cm$^{-2}$.  
Our $K$
= 0 cm$^2$s$^{-1}$ model predicts $N$(CO) = 7.3 $\times$ 10$^{14}$
cm$^{-2}$ at $R$ = 300 AU, considerably less than either observed
value.  The inclusion of diffusion with $K$ = 10$^{18}$ cm$^2$s$^{-1}$
is sufficient to provide a means of keeping molecules in the gas while
using the cool disks that produce better agreement with SEDs, and
without the need to invoke alternative non-thermal desorption
processes.

\begin{deluxetable}{lllllll}
\tablecolumns{7}
\tablewidth{0pt}
%\tabletypesize\scriptsize
\tablecaption{Calculated column densities (cm$^{-2}$) at 1 Myrs at R = 300 AU. 
$a (b)$ represents $a$ $\times$ 10$^b$.  The observed values are
taken from (1) \citet{aikawa02}
(derived from \citet{dutrey97}, (2) \citet{qi01}
(3) \citet{thi04}, (4) \citet{aikawa03}. 
The column densities for LkCa 15 calculated from the single dish 
observations assume that the disk has a radius of 450 AU.  For
TW Hya the disk is assumed to be 165 AU in radius.
% \citet{aikawa02}
%calculate the column densities from the observations of DM Tau
%made by \citet{dutrey97}.  \citet{dutrey97} derive fractional 
%abundances from single disk observations of DM Tau by assuming that
%the average fractional abundance is the same everywhere in the disk
%and assuming a vertical density distribution.  \citet{aikawa02}
%used the density distribution in conjuction with the derived 
%factional abundances to determine an average column density.  
%\citet{qi01} made interferometric observations of LkCa 15.  The
%column densities quoted are beam averaged and can be assumed to
%be valid at $R$ = 300 AU if the emission is resolved and if
%the column densities do not change much across the disk.  \citet{thi04}
%observed the same object but with a single dish telescope and derived
%beam averaged column densities assuming emission is coming from 
%a region within $R$ = 100 AU.  This radial limit is rather arbitrary
%and results in much higher column densities than derived for DM Tau.
%\citet{aikawa02} recalculated the \citet{thi04} data assuming a disk
%size of 375 AU (i.e. the column densities are scaled by (100/375)$^2$)
%which results in values that are much closer to those derived for DM Tau.
%The difference in the derived column densities from the interferometric and single dish 
%data can be attributed to differences in the models used to derive the
%column densities and in the different regions covered by the telescope
%beams.
\label{tab:col}}
\tablehead{
\colhead{Molecule} & \colhead{$K$ = 0} & \colhead{$K$ = 10$^{18}$} & \colhead{DM Tau$^1$} & \multicolumn{2}{c}{LkCa 15} & \colhead{TW Hya$^3$}\\
\colhead{} & \colhead{(cm$^{2}$s$^{-1}$)} & \colhead{(cm$^{2}$s$^{-1}$} & \colhead{} & \colhead{Interferometer$^2$} & \colhead{Single dish$^3$} 
}
\startdata
H$_2$      & 1.1 (23) & 1.1 (23) & \\
CO         & 7.3 (14) & 1.3 (16) & 5.7 (16)       &                    & 1.9 (16)       & 3.2 (16) \\
HCN        & 1.4 (9)  & 6.6 (11) & 2.1 (12)       & 2.4 (13)           & 1.8 (12)       & 9.7 (12) \\
HNC        & 1.1 (10) & 8.8 (11) & 9.1 (11)       & $<$ 5.4 (12)       &                & $<$ 1.4 (12)\\
CN         & 4.4 (11) & 4.0 (13) & 9.5 - 12 (12)  & 9.7 -- 25 (13)     & 1.5 (13)       & 6.6 (13) \\
CH$_3$OH   & 1.9 (5)  & 2.5 (8)  &                & 7.3 -- 18 (14)     & $<$ 7.1 (13)   & $<$ 1.1 (13)\\
H$_2$CO    & 2.9 (10) & 7.1 (12) & 7.6 -- 19 (11) & 7.2 -- 19 (12)$^4$ & 7.1 -- 51 (11) & $<$ 8.0 (11) \\
HCO$^+$    &  5.9 (9) & 2.4 (11) & 4.6 -- 28 (11) & 1.5 (13)           & 3.3 (11)       & 4.4 (12) \\
H$_2$O     & 6.0 (10) & 3.2 (12) &                & \\
C$_2$H     & 1.1 (12) & 7.4 (13) & 4.2 (13)       &                    &              \\
N$_2$H$^+$ & 3.1 (11) & 1.6 (11) & $<$ 7.6 (11)   &                    & $<$ 1.4 (12)   & $<$ 1.0 (13) \\
NH$_3$     & 2.6 (12) & 6.5 (11) &                & \\
\enddata
\end{deluxetable}

\section{Effect of binding energy of CO on the results}

In the previous section we presented the results of models that
used a value of 1210 K for the binding energy of CO \citep{ta87}.  Recent
experimental work by \cite{oberg05} has suggested that $E_b$(CO)
is actually considerably lower and, at 855 K, is close to that of
N$_2$.  This would mean that CO could be more easily returned to the gas phase
by thermal desorption. 
Since the chemistry is very dependent on the efficiency of
thermal desorption, a 
change of this magnitude in the binding energy of a molecule could
greatly change the gas phase chemistry.

To test how this affects our results we ran models at 100 AU and 300
AU using BE(CO) = 855 K. As expected, decreasing the CO binding energy
increases the gas phase column density of CO 
at 100 AU from 1.9 $\times$ 10$^{15}$ cm$^{-2}$ to
2.4 $\times$ 10$^{17}$ cm$^{-2}$ for a model without mixing.  When
mixing with $K$ = 10$^{18}$ cm$^2$s$^{-1}$ is included, the 
change in $E_b$(CO) increase $N$(CO) 
from 4.4 $\times$ 10$^{16}$ to 2.9 $\times$
10$^{17}$ cm$^{-2}$.  The increase is due to the thermal desorption of 
CO from grains, which can now occur at lower temperatures ($\sim$ 17 K), 
resulting in a higher abundance in the midplane.  
At larger radii the change in $E_b$(CO) has less effect as the
temperatures in the midplane are sufficiently cold that thermal
desorption is not so important.  Hence at $R$ = 300 AU
the change in $E_b$(CO) does not affect the CO column density for
$K$ = 0 and only results in an increase of a factor of 2 for the model with 
$K$ = 10$^{18}$ cm$^2$s$^{-1}$.

Most other molecules are
not affected by the presence of extra CO in the gas, but N$_2$H$^+$
shows a drop in column density due its destruction by reaction with CO
in the midplane.  (In models with a higher $E_b$(CO),
where the midplane CO abundance is low, 
N$_2$H$^+$ is destroyed mainly by recombination with electrons).
At $R$ $<$ 100 AU, therefore, N$_2$H$^+$ will not trace the midplane
if the CO binding energy is low, but it will be a possible
tracer of this region at larger radii, where CO is accreted onto 
grains.

\section{Discussion and conclusions}

The inclusion of diffusion can have a considerable effect on model
abundances, with the effect depending on the magnitude
of the diffusion coefficient assumed.  Diffusion tends to smooth out
the vertical abundance gradients and to increase the vertical range
over which molecules are present.  Mixing transports atoms and ions
from the surface photodissociation region into the molecular layer
where they can be incorporated into new molecules.  The transport of
icy grains into warmer regions also affects the abundances of some
(more volatile) molecules and is most effective at $R$ = 100 AU, where
the greatest increase in column density with $K$ is seen.  Our models
show that if $K$ $<$ 10$^{16}$ cm$^2$s$^{-1}$ then diffusion is
unlikely to have much effect, except at small radii ($R$ = 100 AU).
Diffusion does not destroy the three layer chemical structure of the disk
predicted by static models, although the thickness of the molecular
layer does increase.

We find that the inclusion of diffusive mixing can greatly affect the
abundance of many species.  Most molecules show an increase in column
density, but some nitrogen--bearing molecules are decreased.  How the
column density is affected is related to the location of the molecules
in the disk.  Molecules that have their peak abundance at or near the
midplane, e.g.\ N$_2$H$^+$ and NH$_3$, are relatively unaffected by
diffusion, whereas molecules which exist mainly in the molecular layer
show increases in column density.  This increase depends on the
magnitude of $K$, with higher values of $K$ resulting in larger column
densities.

Our calculated midplane ionization fraction is consistent with the
observations of \cite{cec04}, but do not change with the addition of
diffusion.

In conclusion our models show that diffusion can greatly affect
the chemistry in a disk.  It is important to include the 
interplay of chemistry and dynamics in disks since both can
have an effect on the other: the dynamics can change the 
abundance and distribution of many molecules and the chemistry
can affect the ionization fraction, which in turn can determine
whether or not the magneto-rotational instability can 
drive turbulence.

\appendix
\section{\label{ap:spec}The species set}

%\begin{deluxetable}{llllllllll}
%\tablecolumns{10}
%\tablewidth{0pt}
%\tablecaption{The species set}
%\tablehead{}
%\startdata
%\sidehead{Gaseous species}
\[
\begin{array}{llllllllll}
\multicolumn{2}{l}{\mbox{\it Gaseous species}}\\
\mbox{H} & \mbox{H}_2 & \mbox{H$^+$} & \mbox{H$_2^+$} & \mbox{H$_3^+$} & \mbox{He} & \mbox{He$^+$} & \mbox{C} & \mbox{C$^+$} & \mbox{C$_2$}\\
\mbox{C$_2^+$} & \mbox{O} & \mbox{O$^+$} & \mbox{O$_2$} & \mbox{O$_2^+$} & \mbox{N} & \mbox{N$^+$} & \mbox{N$_2$} & \mbox{N$_2^+$} & \mbox{Si} \\
\mbox{Si$^+$} & \mbox{CH} & \mbox{CH$_2$} & \mbox{CH$_3$} & \mbox{CH$_4$} & \mbox{CH$^+$} & \mbox{CH$_2^+$} & \mbox{CH$_3^+$} & \mbox{CH$_4^+$} & \mbox{CH$_5^+$} \\
\mbox{C$_2$H} & \mbox{C$_2$H$^+$} & \mbox{C$_2$H$_2$} & \mbox{C$_2$H$_2^+$} & \mbox{C$_2$H$_3+$} & \mbox{C$_2$H$_4$+} & \mbox{C$_3^+$} & \mbox{C$_3$H$^+$} & \mbox{C$_3$H$_3^+$} & \mbox{CO} \\
\mbox{CO$_2$} & \mbox{CO$^+$} & \mbox{CO$_2^+$} & \mbox{HCO} & \mbox{HCO$^+$} & \mbox{H$_2$CO} & \mbox{H$_2$CO$^+$} & \mbox{H$_3$CO$^+$} & \mbox{HCO$_2^+$} & \mbox{CH$_3$OH} \\
\mbox{CH$_3$OH$_2^+$} & \mbox{OH} & \mbox{H$_2$O} & \mbox{OH$^+$} & \mbox{H$_2$O$^+$} & \mbox{H$_3$O$^+$} & \mbox{NH} & \mbox{NH$_2$} & \mbox{NH$_3$} & \mbox{NH$^+$}\\
\mbox{NH$_2^+$} & \mbox{NH$_3^+$} & \mbox{NH$_4^+$} & \mbox{N$_2$H$^+$} & \mbox{CN} & \mbox{HCN} & \mbox{HNC} & \mbox{CN$^+$} & \mbox{HCN$^+$} & \mbox{HCNH$^+$} \\
\mbox{CNC$^+$} & \mbox{C$_2$N$^+$} & \mbox{H$_2$NC$^+$} & \mbox{HC$_3$N} & \mbox{CH$_3$CN} & \mbox{H$_2$C$_3$N$^+$} & \mbox{H$_4$C$_2$N$^+$} & \mbox{NO } & \mbox{NO$^+$} & \mbox{e$^-$} \\
\\
\multicolumn{2}{l}{\mbox{\it Mantle species}}\\
\mbox{H} & \mbox{H$_2$} & \mbox{C} & \mbox{C$_2$} & \mbox{O} & \mbox{O$_2$} & \mbox{N} & \mbox{N$_2$} & \mbox{Si} & \mbox{CH} \\
\mbox{CH$_2$} & \mbox{CH$_3$} & \mbox{CH$_4$} & \mbox{C$_2$H} & \mbox{C$_2$H$_2$} & \mbox{CO} & \mbox{CO$_2$} & \mbox{HCO} & \mbox{H$_2$CO} & \mbox{CH$_3$OH} \\
\mbox{OH} & \mbox{H$_2$O} & \mbox{NH} & \mbox{NH$_2$} & \mbox{NH$_3$} & \mbox{CN} & \mbox{HCN} & \mbox{HNC} & \mbox{HC$_3$N} & \mbox{CH$_3$CN} \\
\mbox{NO} \\
%\enddata
%\end{deluxetable}
\end{array}
\]
\acknowledgements
This research was conducted at the Jet Propulsion Laboratory,
California Institute of Technology under contract with the
National Aeronautics and Space Administration.  Partial
support was provided by a grant from the NASA Origins Program.

\bibliography{disk}
\end{document}